\documentclass[12pt]{article}
 \usepackage{amsfonts}
 \usepackage{amssymb,amsmath}
 \usepackage{graphicx} \usepackage{epstopdf}
 \newcommand{\crlb}[1]{\label{#1}\\[2pt]}
 
 \newcommand{\crld}[1]{\label{#1}}
 \newcommand{\eela}[1]{\quad\hbox{\scriptsize{#1}}\label{#1}\end{eqnarray}}
 \newcommand{\eelb}[1]{\label{#1}\end{eqnarray}}
 
 \newcommand{\newsecb}[2]{\section{#1}\label{#2}\setcounter{equation}{0}}

 \newcommand{\nolabels} {\def\eel{\eelb}\def\eeql{\eeqlb}  \def\crl{\crlb} 
 \def\newsecl{\newsecb}\def\bibiteml{\bibitem} \def\citel{\cite}\def\labell{\crld}}
\newcommand{\eeqla}[1]{\quad\hbox{\scriptsize{#1}}\label{#1}\end{aligned}\end{equation}}
\newcommand{\eeqlb}[1]{\label{#1}\end{aligned}\end{equation}}

\newcommand\publishversion  {\nolabels\setlength{\textheight}{8.38in}\setlength
    {\oddsidemargin}{0in} \setlength{\textwidth}{6.2in}\setlength{\topmargin}{-0.2in}}

\def\beq{\begin{equation}\begin{aligned}}		\def\eeq{\end{aligned}\end{equation}}
\def\be{\begin{eqnarray}}  					\def\ee{\end{eqnarray}}		
   \def\bi#1{\begin{itemize}\item[#1]} 	      	   \def\ei{\end{itemize}} 
   \def\eqn#1{(\ref{#1})}
\def\Tr{{\mbox{Tr}}\,}   	 \def\fn{\footnote}	\def\nm{\nonumber} 

		 \def\del{\delta}  

 \def\bet{\beta}  \def\gam{\gamma}      
 
 \def\del{\delta}      \def\Del{\Delta}    
       \def\lam{\lambda}      
             \def\vv{\varphi}    
             \def\rr{\varrho}       
      \def\tht{\theta}  
               
       \def\W{\Omega}

     \def\OO{{\mathcal O}}  
  \def\ZZ{\mathbb{Z}}   
 \def\pa{\partial} \def\ra{\rightarrow} 
  
 \def\dd{{\rm d}}  \def\bra{\langle}   \def\ket{\rangle}

\def\fract#1#2{{\textstyle\frac{#1}{#2}}}	 	 	
\def\ffract#1#2{\raise .2 em\hbox{$\scriptstyle#1\,$}\kern-.34 em/\kern-.34 em\lower .15 em \hbox{$\scriptstyle\,#2$}}
					
\def\tl#1{\tilde{#1}} 
\def\ex#1{e^{\textstyle#1}} 			
\def\bpmatrix{\begin{pmatrix}} 			\def\epmatrix{\end{pmatrix}}
\def\bmatrix{\begin{matrix}} 			\def\ematrix{\end{matrix}} 
\def\bcenter{\begin{center}}			\def\ecenter{\end{center}}

\def\lowerheightgth#1#2#3{\(\raise-#1\hbox{\includegraphics[height=#2]{#3}}\)}
\def\lowerwidthgth#1#2#3{\(\raise-#1\hbox{\includegraphics[width=#2]{#3}}\)}

  \def\Planck{{\mathrm{Planck}}}

  \def\BH{{\mathrm{BH}}}

\def\weglaten#1{}	
 
    \def\Ret{\\[10pt]}
	
  \publishversion
\begin{document}
\begin{titlepage} 
\title{
How an exact discrete symmetry can
preserve black hole information\fn{Presented at DICE2022, Castiglioncello, Italy, May 19-22, 2022.}\\
\normalsize or\\
\large Turning a Black Hole Inside Out
\author{Gerard 't~Hooft}}
\date{\normalsize
Faculty of Science,
Department of Physics\\
Institute for Theoretical Physics\\
Princetonplein 5,
3584 CC Utrecht \\
\underline{The Netherlands} 
http://www.staff.science.uu.nl/\~{}hooft101}
 \maketitle

\begin{quotation} \noindent {\large\bf Abstract } \\[10pt]
	To apply the laws of General Relativity to quantum black holes, one first needs to remove the horizon singularity
by means of Kruskal-Szekeres coordinates. This however doubles spacetime, which thereby is equipped with an exact binary symmetry. All particles near a black hole share the same symmetry, and conservation of this
symmetry may completely remove the information paradox: the quantum black hole has no interior, or equivalently,
the black hole interior is a quantum clone of the exterior region. These observations, totally overlooked in most of the literature on quantum black holes, resolve some issues concerning conservation of information.
Some other problems do remain .
 
 \end{quotation}\end{titlepage}

\newsecl{Introduction}{intro.sec} \setcounter{page}{2}
A silent revolution took place in the author's understanding of the quantum mechanical equations for black holes, as first mentioned in Ref.\,\cite{GtHclones.ref}. In summary, what happens is that the space-time singularity associated to the horizon of a black hole, can easily be removed by an appropriate coordinate transformation, showing that a black hole horizon is about as regular as the apparent singularity of planet Earth at its North and South poles: it is a coordinate singularity. Removing this singularity should answer all questions about the nature of space and time there, but it leads to a new question instead. 
Apparently, beyond the coordinate singularity, space and time are extended into a new copy of the existing space-time. Our new insight, about which we also reported in Ref.\,\cite{GtHPadova.ref}, is that this is not a new space-time region at all, but merely an exact mirror image of existing space-time. The black hole carries along a clone, not only of itself, but also of the entire surrounding universe. 

An observer\fn{The concept of an observer can perhaps better be replaced by the concept of observable variables, so that some problems such as finite-size effects and finiteness of memory space, are more easy to avoid, but clearly, utmost diligence would be asked for.}, moving from ordinary space-time to the cloned region will notice no differences at all, but in fact space-time in this region is quite remarkable, because the clone is connected to the original \emph{via} time reversal. Clearly, this is not an ordinary symmetry, but it does allow us to handle it as such. We can limit ourselves to states in Hilbert space that are either symmetric or anti-symmetric\fn{symmetric for the real parts of wave functions, and antisymmetric for the imaginary parts, as will be explained, see section \ref{unitary.sec}.}  under the switch between the black hole and its clone. 
As we shall show briefly, this reduction has a big effect on the expected properties of the black hole. Its entropy is now only half of what is usually found, and consequently, the temperature of the emitted radiation doubles.

But we shall come to that; let us first return to some basic physical issues concerning quantum gravity. Usually, our starting point is chosen to be Newton's force law,
	\be F=\frac{G\,m_1\,m_2}{r^2}\ , \eel{Newton.eq}
where one imposes the condition that this law needs to be made compatible with the highly successful laws of special relativity.
As is well-known, one then ends up with General Relativity. It is found that space and time are curved, and the curvature is everywhere in the vicinity of the source. Making the next step, needed to impose the laws of quantum mechanics, is now very hard, and usually comes with new assumptions and unprovable principles, such as string theory and holography. There is nothing wrong with such approaches, except that the results are generally uncertain. One has to understand Nature's (or God's\fn{I should henceforth avoid using God's name, as this apparently suggests religious feelings that I do not have.}) way; you have to be religious to understand that, and if you are not religious, you just miss the boat.
	
	An approach that  requires (almost\fn{A critical reader may observe that we do make use of a belief: the fact that black holes should in no way differ from ordinary particles in the way they obey quantum mechanical laws. Yes, I admit that this is a belief.}) no religion is the path offered by black holes. If we remove the coordinate singularities by a coordinate transformation, we only encounter small amounts of local curvature. However, the effects of gravity still diverge in a remarkable way, which is the effects of the gravitational force acting between in- and out-going matter. This diverges badly, in a process that explodes exponentially in time, but in a way that triggered our interest: this divergence \emph{only} occurs on the horizon. Elsewhere in space-time, one may assume that standard knowledge of general relativity and quantum mechanics offers sufficient clues as to what happens there, so, \emph{unintelligible curvature} does not occur everywhere, and we may attempt to couple understandable features of regions of space and time, while reserving our attention to the tiny subset of space-time points right at the horizon. This is where a black is connected to  its clone at the other side.
	
	Instead of Newton's law, Eq.~\eqn{Newton.eq} , we focus on the Shapiro effect.
	 This effect brings about gravitational lensing, as is well known in astronomy:
	\be \del u^-=-8\pi G\,p^-\,\log|\tl x_1-\tl x_2|\ , \eel{Shapiro.eq}
where \(p^-\) is the momentum of the source, which we take as a vector in the light cone 
minus-direction, while \(u^-\) is the displacement it generates in a reference mass, also
 in the minus-direction. \(\tl x_1\) and \(\tl x_2\) are the transverse components of the locations of source and effect.
The most noticeable distinction with Newton's law is the dependence on these coordinates. 
It goes as \(1/r^2\) in the Newtonian case, while it is only logarithmic in the case of Shapiro. 
The logarithm reflects the fact that the associated phenomenon takes place in a two-dimensional 
subspace of space-time. Laplace's field equation generates logarithms in two dimensions.
	
The Shapiro equation \eqn{Shapiro.eq} offers an easy way to handle the gravitational interactions between in- and out-particles. One finds that, in the reference frame of the out-particles, the momentum \(p^-\) of the in-particles increases exponentially as a function of the  time	separation between in and out. Remarkably, here \emph{momenta} of the in-particles affect the \emph{positions} of the out-particles. By applying Fourier transformations, one derives directly that (apart from a sign) the same relation holds if for both particles momenta and positions are interchanged. This is time reflection symmetry, see Section~\ref{Shapiro.sec}. Readers familiar with these issues might want to proceed directly to Section~\ref{unitarity.sec}, where some subtle novelties enter.
	
\newsecl{Perturbations around the stationary black hole}{stationary.sec}
	
Many researchers now continue with the black hole as it is being formed by some implosion event.\cite{Ashtekar.ref}
 It then seems as if the details of this implosion are intimately responsible for what happens next. We claim however that this only holds for time spans shorter than the \emph{scrambling time.} 
This is the time it takes for a particle to reach the horizon within a few Planck lengths, in natural units:
\be T_S=\OO\big(M_\BH\log(M_\BH/M_\Planck)\big)\ . \eel{scramblingtime.eq} 
Classically we have the no-hair theorem stating that the details of the implosion event are 
 immaterial. In practice one expects that only minute quantum details will continue to depend on 
 details of the original implosion, but this situation would be similar to what happens to an atom 
 after its formation at the beginning of the universe. Surely such details will determine which of its 
 quantum states it will be in right now, but as physicists we should be primarily interested in a proper 
 categorisation of all possible quantum states, and the Schr\"odinger equation dictating how these
  evolve one into another. We here take the same attitude regarding black holes. Its history before 
  scrambling time, or its evaporation process after scrambling time,  should not be relevant for its 
  physical equations at present. 

Soon after its formation, the black hole is almost stable, absorbing and emitting particles every now and then, 
according to a Schr\"odinger equation and a Hamiltonian connecting all microstates. It is our job to use basic 
principles for deriving the proper classification of states and the fundamental Hamiltonian. At some decent 
distance from the horizon, the categorisation is clear and the Hamiltonian is known, it is called the Standard 
Model of the subatomic particles, which can be put in a gently curved space-time.
 At all space-time points where  the temperature dropped to below a 
few TeV, we know how it generates the answers to our questions.
\def\hokje{\(\,\)}

We now assume that the complete set of states is determined by the spectrum of particles excited thermally 
at all distances beyond a few Planck lengths from the horizon, where the exact numbers are still to be derived.	
Since the particles are thermal and the local temperature is expected to have dropped well below the Planck 
temperature, the effects of these particles on the black hole space-time metric must have dropped to values 
small enough to neglect them for the time being. We plan to postpone these details to being  
justified a posteriori (See Section \ref{conc.sec}). Right now we can observe that the sets of states and operators used at time scales 
that may vary over more than one unit of the scrambling time will definitely not commute with one another. 
	The Schwarzschild metric, in Schwarzschild's coordinates, demonstrates clearly the time dilation invariance:
	\be \dd s^2&=&\frac 1{1-\fract{2GM}r}\dd r^2-\big(1-\fract{2GM}r\big)\dd t^2+r^2\dd\W^2\ , \labell{Schw.eq}\\
	\hbox{where\qquad}			\W&\equiv&(\tht,\,\vv)\ ,\nm\Ret
			\dd\W&\equiv&(\dd\tht,\,\sin\tht\,\dd\vv)\ .\eel{Schwarz.eq}
  It is sketched in Figure~\ref{Schw.fig}a.
  The regular coordinate frame is the Kruskal\,\cite{Kruskal.ref} Szekeres\,\cite{Szekeres.ref} coordinates, 
  to be referred to as tortoise coordinates \(x\) and \(y\), see Figure \ref{Schw.fig}b.
  One defines
  	\be x\,y&=&\Big(\frac r{2GM}-1\Big)\,\ex{r/2GM}\,;\labell{tortoise1.eq}\\[4pt]
	y/x&=&\ex{t/2GM}\ , \eel{tortoise2.eq}
in terms of which, the spacetime described by Eq.~\eqn{Schw.eq} is regular at the horizon:  
  	\be \dd s^2\ =\  \frac{32(GM)^3}r\,\ex{-r/2GM}\dd x\,\dd y\ +\ r^2\dd\W^2\ .  \eel{KSmetric.eq}

  \begin{figure} \begin{center} 
		\vskip 20pt time \(\uparrow\) \hskip-10pt 
		\lowerwidthgth{90pt}{150pt}{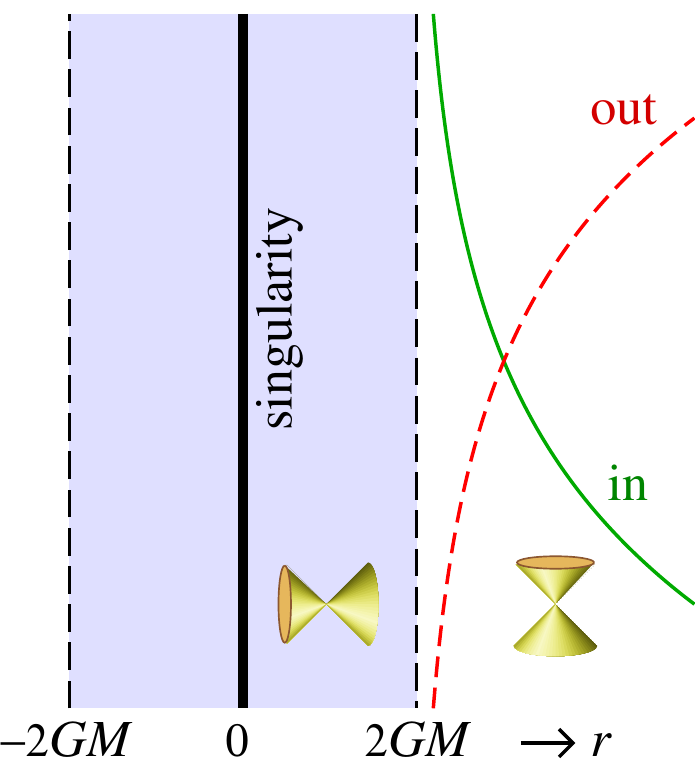} \qquad \quad\lowerwidthgth{80pt}
		{150pt}{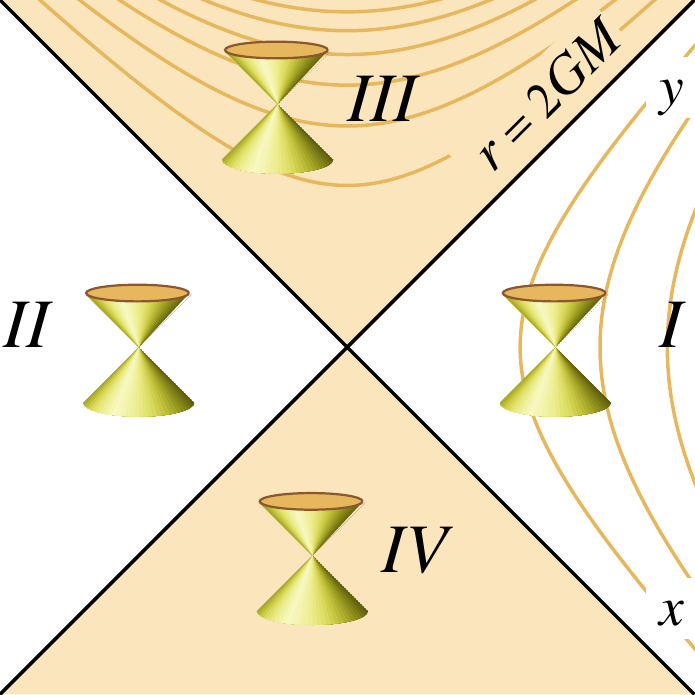}  \\
		\hskip-100pt a) \hskip 230 pt b)\end{center} \vskip-20pt
		\begin{caption} {a)	The Schwarzschild metric, in Schwarzschild coordinates.
		Vertical dashed lines: the event horizon. 
		Curved lines: light like radial geodesics: \(y\,=\,\)const (smooth), and \(x\,=\,\)const (dashed). 
		Cones: the orientation of the local light cones. 
		Angular coordinates \((\tht,\,\vv)\) are not shown. \quad
		b) The same space-time in Tortoise coordinates \((x,\,y)\), showing regions \(I - IV\) and the orientations of the
		 local light cones. Curves show \(r=\)\,const. lines.}\labell{Schw.fig} 
		 \end{caption}
	\end{figure}
	
All of space-time as we know it, is covered by the set \(\{x\ge 0\,,\ y\ge 0\}\) (region \(I\)).
As we mentioned in the Introduction, this new coordinate frame   allows for an analytic  continuation towards another black hole space-time, defined by the new region \(II\): \ \(\{x\le 0\,,\ y\le 0\}\). In contrast, the regions \(III\) where \(x\le 0\) and \(y\ge 0\), and   \(IV\) where \(x\ge0,\ y\le 0\), are situated beyond the infinite future or infinite past, and these are physically less bothering. 
	
Both coordinate frames depicted in Figure \ref{Schw.fig} are of importance, and we have to understand better how they are related.  The \emph{arrow of time} is oriented upward in all outside regions of the Schwarzschild frame Fig. \ref{Schw.fig}a. In its inside regions it is pointing inwards, and this means that a local observer will be grilled at the singularity soon after having passed the horizon. But we are more interested in the way regions \(I\) and \(II\) are linked together, in Fig. \ref{Schw.fig}b.

The arrow of time can be read off from Eq.~\eqn{tortoise2.eq}. It implies that, as a local observer in the horizon region sees time always moving in the upwards direction inside its local light cone, an outside observer using the growing time coordinate \(t\), will be moving downwards in region \(II\). Wave functions in \( I\) will rotate as \(\ex{-iEt}\), but in region 
\(II\) as \(\ex{+iEt}\). However, as soon as the Standard Model applies, such negative energy modes are forbidden. 

Either one can say that the Dirac ket states seen by the local observer relate to the  bra states of the global observer, or one can say that, in the static black hole approximation, the states seen by the local observer in region \(II\) describe a universe that is \emph{almost full} rather than almost empty. Considering the fact that, as derived by Hawking\,\cite{Hawking.ref}, quantum effects generate particles at the crossing points of the two horizons, which are seen as particles in \emph{both} regions, the idea that we have a space full of particles in region \(II\) leads to a satisfactory image of a quantum mechanical leak at that point, producing a steady shower of Hawking particles.

At first sight one might object that there is no such leak; a time translation for the outside observer leads to a  Lorentz transformation for the local observer, and it might seem that no particles should leak out. Now of course quantum effects may generate tunnelling, and there is a more calculable genuine effect that generates such tunnelling: the Shapiro effect mentioned in the Introduction.

 \newsecl{The Shapiro effect}{Shapiro.sec}
 	How to employ the Shapiro effect\,\cite{Shapiro.ref} in understanding the dynamics of the Hawking effect, was elaborately explained in our previous publications\,\cite{GtHclones.ref, GtHFirewall.ref}.	Eq.~\eqn{Shapiro.eq} is derived by first considering the Schwarzschild metric of a particle at rest, and then subjecting that to a Lorentz transformation with a large \(\gam\) factor. The particle momentum is the limit \(p^-\ra \gam m\) for large \(\gam\), and small rest mass \(m\). The result\,\cite{Bonnor.ref, AichelbSexl.ref} still describes flat space-time, but General Relativity tells us that there is curvature, delta-peaked at the location of the shock wave associated to the fast source particle. Next, one generalises the equation for the case that the source mass goes inwards through an \(S_2\) surface, the horizon, rather than a flat shock wave. The result\,\cite{GtHDray.ref} is almost the same, but the transverse separation \(\tl x-\tl x'\) is now described by an angular separation, the logarithm turns into a Laplace function on the sphere. We see that Eq.~\eqn{Shapiro.eq} turns into
	\be \del u^-=Gp^- f(\W,\W')\quad\hbox{\ with\ }\quad (1-\Del_\W)f(\W,\W')=8\pi G\del^2(\W,\W')\ . 
	\eel{sphericalGreen.eq}
Here, \(\W\) is the solid angle at which the out-particle leaves, and \(\W'\) is where the in-particle entered the black hole horizon.

We add normalisation constants that turn the tortoise coordinate \(x\) into the lightcone coordinate \(u^-\) and \(y\) into \(u^+\). For large black holes, the metric in the \(u^\pm\) coordinates is almost flat. Henceforth, we ignore the slight curvature that remains.

With \(u^\pm\) we indicate the positions at the future or past  horizon, \emph{i.e.}, the time at which the particles enter or leave. \(p^\pm\) are their momenta.	
These equations are linear in the position and momentum variables, and consequently, it is easy to generalise this result to handle the case of many particles going in or out. These lead to momentum and position distributions on the horizons:
	\be p^\pm\ra p^\pm(\W)\,,\qquad u^\pm\ra u^\pm(\W)\ . \eel{momposdistr.eq}	
And this allows us to use  spherical wave expansions\fn{These distribution functions are to be defined accurately: the momenta emerge as sums of \(\delta\) distributions, while positions are represented as centre of mass positions. See  for example Ref.\,\cite{GtHPadova.ref}, sections 3 and 4.}
	\be u^\pm(\W)=\sum_{\ell,m}u^\pm _{\ell\,m}Y_{\ell\, m}(\W)\ , \qquad 
	p^\pm(\W)=\sum_{\ell,m}p^\pm_{\ell\,m}Y_{\ell\, m}(\W)\ .	\eel{harmexp.eq}
Since, for each spherical harmonic, \(\Del=-\ell(\ell+1)\), we can write Eq.~\eqn{sphericalGreen.eq} as	
	\be\del u^-_{\ell m}=\frac{8\pi G}{\ell^2+\ell+1}p^-_{\ell m}\ . \ee
Now we can use the commutator equations for all in- and out-going particles\ \(i,\,j\),
	\be {}[u_i^\pm, \ p_j^{\mp}]=i\del_{i\,j}\ ,\ee
to arrive at the algebra
	\be {}[u^+_{\ell\,m},\ u^-_{\ell'\,m'}]=i\lam_\ell \,\del_{mm'}\,\del_{\ell \ell'}\ ;\qquad
	\lam_\ell=\frac{8\pi G}{\ell^2+\ell+1}\ .\eel{ellmalgebra.eq}

A few remarks:\\ We dropped the symbol \(\del\) from \(\del u^\pm\), which means that it is assumed that all displacements \(\del u\) added up give the positions \(u\) themselves, provided coordinates are chosen appropriately. The same holds for the momenta. 

Secondly, all different \((\ell,\,m)\) modes decouple, so we obtain new, one-dimensional quantum systems that generate the total Hilbert space in a very simple manner. These equations are highly trivial.

Furthermore, the Fourier transformation from momenta to positions and back is a \emph{unitary} transformation. Hence, the algebra \eqn{ellmalgebra.eq} should act as a perfect, information preserving, mapping from in- to out-particles (and back).

\newsecl{Restoring unitarity}{unitary.sec}

But then there is a more serious problem that has to be addressed. The Fourier transforms arrived at in the previous section, are only unitary if the data stretch over the entire real line both for the \(u^+\) and the \(u^-\) variables. But the physical data in the observable part of the universe are exclusively on the half-lines \(u^+>0\) and \(u^->0\). If we drop the data on the negative half-lines, the Fourier transformation is not unitary at all.

The cure to this problem may well come from the observation proposed in the Introduction section: just choose region \(II\) to be a quantum clone of region \(I\). In that case, also the Fourier variables in region 
\(II\) are quantum clones of those in region \(I\). Is this an elegant solution to our problem?

Unfortunately no. If all wave functions were real, then indeed the equation 
	\(\psi(u)=\psi(-u)\)
copies correctly in Fourier space as well: \(\hat\psi(p)=\hat\psi(-p)\). But wave functions are not in general real. Can we impose reality as an extra condition? 

In ordinary quantum mechanics, the condition that wave functions are real is not impossible. Taking the Hamiltonian to be an imaginary, antisymmetric matrix would be allowed in quantum mechanics, provided that we accept the corresponding symmetry in the energy spectrum:   every state with energy \(E\) in the energy spectrum, is associated to a state with the opposite sign of the energy. Reality of the wave function implies a symmetry	
\be E\leftrightarrow -E\eel{Esymm.eq} 
 in the energy spectrum.
Remember that, right at the beginning of our treatment, it was imposed that for ordinary particles the effect of the particles to the total mass of the black hole may be ignored, so indeed, we may assume the condition that the entire energy spectrum of particles is shifted to obey the symmetry \eqn{Esymm.eq}.

This line of arguments needs to be carefully checked to ascertain that it be realistic. 
	 
	 In previous work, it was proposed to combine the transformation \(u\leftrightarrow -u\) with the \emph{antipodal mapping}:
	 \be \W=(\tht,\,\vv) \leftrightarrow -\W\equiv (\pi-\tht, \vv+\pi)\ ,\ee
	but this was found to lead to other difficulties: the antipodal mapping is the mapping that replaces all spherical harmonic functions \(Y_{\ell\,m} \) by \((-1)^\ell\,Y_{\ell\,m}\). But since we already had that the mapping \(I\leftrightarrow II\) corresponds to \(u^\pm\leftrightarrow -u^\pm\)\,, this would imply
	 \be u^\pm_{\ell,m}(\W)=(-1)^{\ell+1} u^\pm_{\ell,m}(\W)\,, \ee
	so that \(u^\pm=0\) if \(\ell\) is even, which cannot be squared with the commutation equation saying that all \(u\) variables obey \([u^+,\,u^-]=i\lam_\ell \), see Eq.\,\eqn{ellmalgebra.eq}. Therefore the transition to antipodes is herewith dismissed.
	
	An other possibility was proposed im 1984 by the author\,\cite{GtHambiguity.ref}: should we regard all states in region \(II\) as the bra states \(\bra\psi|\) associated to the ket states \(|\psi\ket\) in region \(I\)? The elegance of this proposal is that, together, these structures form the density matrices of the system in region \(I\). Is it possible to accept the mathematics of the states in region \(I\) and \(II\) taken together, as representing the quantum density states? Actually, this proposal almost coincides with our attempt to impose the symmetry \eqn{Esymm.eq}. It is not as yet ruled out, and it seems to be the only possibility left. Note that we are forced to supplant our complex wave functions by real ones. It is as if the use of complex  global \(U(1)\) symmetries is forbidden, which might enforce violation of exact, additive, infinite conservation laws. These are replaced by a \emph{discrete \(\ZZ_2\) symmetry} that switches the signs of the tortoise coordinates \(x\) and \(y\).
	
And it does lead to a more general, really important observation: this proposal replaces Hawking's calculation for the temperature of a black hole by another one!
	
If we use density matrices to compute probabilities, we find that probabilities are proportional to the matrix elements of the density matrix, whereas Hawking regards these as quantum vectorstates, and consequently, he averages over the squares of the states describing region \(II\) as these can't be observed. 

Thus, in the thermal Boltzmann factors, we now disagree.  In our expressions, the temperature of radiation emitted by a black hole, of a given mass,  must be twice the temperature found by Hawking.

	\newsecl{Black hole thermodynamics in a nut shell}{thermo.sec}
In statistical physics and quantum field theory it is well-known that a unitary evolution operator, \(\ex{-iHt}\) can be analytically continued towards imaginary time \(t\ra -i\bet\) to become the operator \(\ex{-\bet E}\), which is exactly the operator that produces a thermal probability distribution if the temperature \(T\) obeys 
	\(\bet=1/kT\ ,\) with \(k\) being the Boltzmann constant.\,\cite{Sym1.ref,Sym2.ref}.
One also finds that the distribution of the energy levels is given by the entropy \(S(E)\) if we enumerate the energy eigenstates. In a grand canonical ensemble, one would write \(\rr(E)\dd E\) for the number of energy levels between \(E\) and \(E+\dd E\), to define the free energy \(F(\bet)\) by
\be Z(\bet)=\int_0^\infty \rr(E)\dd E\,\ex{-\bet E}=\int_0^\infty \ex{S(E)-\bet E}\dd E=\ex{-\bet F}\ ,\eel{grand.eq}
where \(S(E)=\log(\rr(E)\) is the entropy of the system if \(E\) is handled as the expectation value \(\bra E\ket\).
The usual thermodynamical equations emerge if we assume \(\bra E\ket\) to be  the macroscopic energy, \(F\) is the free energy, and we use 
\be \bra E\ket =-\frac\pa{\pa\bet} \log Z(\bet)\ . \ee

In black holes however, one expects the integral \eqn{grand.eq} to diverge, as the entropy increases as \(E^2\).
This leads to a negative specific heat , which destabilises the black hole, and therefore it is better to go to a micro-canonical ensemble, where we keep the total energy fixed. In a micro-canonical sample, we define the entropy by ordering all energy eigenstates, writing them as \(E(n)\), and we focus on a single value for \(n\). For large systems, such as large black holes, one can take the continuum limit. Then we write
\be \ex{S(E)}=\frac{\dd   n}{\dd E}\ ,\qquad\frac{\dd S(E)}{\dd E}=\bet(E)\ .\eel{expS.eq}
Here, one defines \be\log Z= -\bet F =	E-S/\bet=E-TS\ \ee
(in units where \(k=1\)). While in the grand canonical ensemble, the temperature could be chosen, here in the micro-canonical ensemble,  \(\bet\) is a fixed function of the black hole mass, being the total energy \(E\).	
	
The fact that \(\bet\) is fixed, originates from the fixed background configuration, being the Schwarzschild metric. 

What is the number \(\bet\) for a black hole with mass \(M\)? Consider all states on a trajectory in complex space-time. Choose the trajectory in tortoise coordinates to be	
\be x=A\,\ex{i\vv}\ ,\quad y=A\,\ex{-i\vv}\ ,\quad xy=|x|^2=A^2\quad \hbox{is fixed}\ ,\quad t=-4iGM\vv\ . 
\eel{xypath.eq}
	
We are dealing with a quantum system where all states in the stretch at imaginary time \(t\) ranging from zero to \(-i\bet\) can be excited to arbitrary values. The total number of energy eigenstates is then given by
 \(\ex{S(E)}=\Tr\big(U(-i\bet)\big)\), where \(\bet\) is the total length of the path. The length of this path is fixed if we assume the end point of the path is the first point where \(x\) and \(y\) take on the values they had at the starting point. This is at \(\vv=2\pi\) in Eq.~\eqn{xypath.eq}. One thus finds that
 \be kT=1/\bet=1/8\pi GM\ ,\eel{HawkT.eq}
 which is \(kT=\fract{\hbar c^3}{8\pi G\,M_\BH}\) when units are put back in. This is Hawking's famous result\,\cite{Hawking.ref}.

Now consider the theory where all states in region \(II\) are   determined by a mapping from region \(I\)\,,
where the exact nature of the mapping is immaterial; it just simply must be a unique	mapping.
Let us again follow our trajectory  \eqn{xypath.eq} in imaginary time. 	At \(\vv=\pi\) we are at \(t=-4\pi GM\) and we arrive at the spot where  \((x,y)\) take the values \((-x,-y)\), which is where the clone of the starting point begins.   Therefore, the trajectory closes at \(\vv=\pi\), which is at time \(t=-4\pi iGM\)\,, so that \(\bet=4\pi GM\ ,\) and the temperature turns into 
	\be kT=\frac{\hbar c^3}{4\pi G M_\BH}\ .\eel{newtemp.eq}
Inspecting Eq. \eqn{expS.eq}, we find that not only \(\bet\) but also the total entropy takes on half the value of Hawking's calculation, Eq.~\eqn{HawkT.eq}. We can understand why this is so. In terms of our local theory, the entropy is an integral over Euclidean spacetime. It counts all independent quantum states in space-time, \emph{but the states localised after the  point \((-x,-y)\) are not independent,} they are the states in region \(II\), which are the same states as the ones seen before in region \(I\). They should therefore not be counted again.

\newsecl{Conclusions}   {conc.sec} 

Our revision of the black hole temperature, Eq.~\eqn{newtemp.eq}, is a direct consequence of the fact that we identify the states in the second region of the Penrose diagram with the ones in region \(I\). We do not merely claim that the particles emerging there, carry all information of the in-going particles, they \emph{are} the in-going particles. This is easiest to comprehend if we return to the Schwarzschild coordinates. There,  one can argue that the cusp singularity at the horizon is somewhat more delicate than just a coordinate artefact. It asks us not to consider all states you get by filling up all of space-time there with any states we like, but we have to fill in region \(II\) exactly as region \(I\). Here, the black hole we describe, fundamentally differs from Rindler space, Rindler space is not just a big black hole because, from the outset, it contains different regions \(I\) and \(II\). One can't create such configurations from the Schwarzschild metric; this is where the Kruskal-Szekeres coordinates for a black hole  are somewhat deceptive, suggesting a world that does not exist. It does exist in Rindler spacetime.

Note that whenever we throw something in, it is seen as in going material both in region \(I\) and in region \(II\).
For instance, when we throw in a dust shell of matter, both regions \(I\) and \(II\) undergo modification. After a large amount of time (more than the scrambling time), the dust shell should become invisible. This only happens if a dust shell is assumed to enter also in region \(II\). This does not only hold for dust shells but for all events in the outside world.

At the beginning of our presentation, in Section~\ref{stationary.sec}, we mentioned some a posteriori verifications.  In particular we claim that a stationary black hole is the appropriate background for an accurate description of  quantum black hole dynamics, as soon as it is large enough compared to the Planck scale. To really appreciate our procedures, our paper was not enough; one should do much more test calculations to appreciate how our picture hangs together. We do hope that our general philosophy will be receiving more attention than it had so-far. The fact that the Hawking temperature is modified by a factor 2 indicates that more is going on than merely a change in semantics.

 \end{document}